\documentstyle[aps,prl]{revtex}

\input epsf

\begin{document}

\draft


\title{Localized moving breathers in a 2-D hexagonal lattice}

\author{J. L. Mar\'{\i}n\thanks{marin@wanda.unizar.es}, J. C.
  Eilbeck\thanks{chris@ma.hw.ac.uk}, and F. M. Russell}

\address{Department of Mathematics, Heriot-Watt University,\\
  Edinburgh EH14 4AS, U.K.}

\date{\today}

\maketitle

\begin{abstract}
  We show for the first time that highly localized in-plane breathers
  can propagate in specific directions with minimal lateral spreading
  in a model 2-D hexagonal non-linear lattice.  The lattice is subject
  to an on-site potential in addition to longitudinal nonlinear
  inter-particle interactions. This study investigates the prediction
  that stable breather-like solitons could be formed as a result of
  energetic scattering events in a given layered crystal and would
  propagate in atomic-chain directions in certain atomic planes. This
  prediction arose from a long-term study of previously unexplained
  dark lines in natural crystals of muscovite mica.
\end{abstract}



\narrowtext

The response of a nonlinear 2D atomic lattice when embedded in a
surrounding 3D lattice is presently of much interest in connection
with transport phenomena in layered crystals.  Two examples of such
crystals are the copper-oxide based high temperature superconductors
and the potassium based silicate muscovite mica.  Fortunately the
optical transparency of the latter allows the study of energetic
events at the atomic level. This is possible in mica because of tracks
made visible by impurities of transient defects, through triggered
solid-state phase transitions.  A study of these tracks arising from
nuclear scattering events led to the suggestion that energy could be
transported over large distances through the crystal by some sort of
energetic intrinsic localized mode (ILM) on the lattice~\cite{srv93}.  The
interesting and novel aspect of this prediction was that these ILMs
would travel along the crystal axes and remain localized in both
longitudinal and transverse directions with little or no lateral
spreading.  This is despite the fact that the 2D lattice has full
hexagonal symmetry.  This one-dimensional behavior in a
two-dimensional lattice was called quasi-one-dimensional (QOD) and the
resulting ILMs were called ``quodons''~\cite{rc95}.

The purpose of this present letter is to demonstrate numerical
simulations of a mica-like model which support the QOD conjecture.
Our intention is not to provide a detailed model of the mica system at
this stage, but merely to present a simplified model which maintains
much of the overall qualitative features of the mica system and yet
can be easily studied.  The model is general enough to suggest the
possibility of QOD effects in other hexagonal crystal structures, and
even other geometries such as square lattices.

\begin{figure}
\epsfxsize=0.45\textwidth\epsfbox{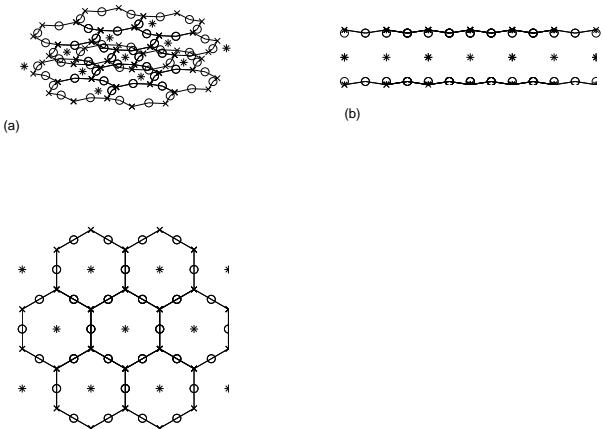}
\caption{The layered structure of muscovite mica. Oxygen atoms are
  marked by $\circ$, the silicon atoms by $\times$, and the
  potassium atoms by $*$.}
\label{fig:mica}
\end{figure}

Muscovite mica has a layered structure.  A conspicuous feature of this
structure is the mono-atomic planes of potassium atoms with large
interatomic spacing, tightly sandwiched between planes containing
oxygen atoms, the oxygen atoms bound in compact silicate layers which
impose a hexagonal structure.  The silicon atoms in these silicate
layers also lie in planes but are further away from the potassium
planes. These features are indicated in Fig.~\ref{fig:mica}, which
shows the symmetry about the potassium plane.  We are interested here
only in the dynamics of the potassium atoms when moving in their 2D
hexagonal plane as a result of in-plane impulses.  The motion of the
potassium atoms about their equilibrium positions is influenced
strongly by the adjacent planes of oxygen atoms, which form a nearly
rigid lattice and therefore give rise to strong on-site potentials.

We assume the potassium atoms are subject to a nonlinear coupling with
the other potassium atoms, and to a nonlinear force due to the fixed
oxygen atoms in the planes above and below.  The (classical)
Hamiltonian of the system is:
\begin{equation}\label{eq:hamilt}
  H= \sum_{i,j} \left[\frac12 \dot{\vec{u}}_{i,j}^2 + \alpha
    V(\vec{u}_{i,j}) + \frac12 \lambda \sum_{i',j'} W(\vec{u}_{i,j},
    \vec{u}_{i',j'}) \right]
\end{equation}
where the indexes $(i,j)$ represent the coordinates of the atomic
(equilibrium) positions in terms of a lattice basis of vectors
$(\vec{a},\vec{b})$. We will denote the lattice spacing by
$a=|\vec{a}|=|\vec{b}|$. The vectors $\vec{u}$ represent
\emph{relative} displacements from these positions. The primed
quantities represent the interacting neighbors for each site, and the
$\frac12$ factor is there to avoid double counting interactions.  In
our case, we have restricted our simulations to nearest neighbor
interactions.  We use radial potentials for both the on-site and the
K-K potentials.

For the interatomic K-K potential we use
\[
W(\vec{u},\vec{u}') = W(|\vec{u}-\vec{u}'-\vec{c}|)= W(|\vec{d}|)
\] 
where $\vec{c}$ is the equilibrium relative position vector, that is,
a lattice vector with $|\vec{c}|=a$.
$\vec{d}=\vec{u}-\vec{u}'-\vec{c}$ is the actual separation vector
between the $(i,j)$ and the $(i',j')$-th sites.  The functional form
of $W$ is chosen to be a classical 6-12 Lennard-Jones potential
\begin{equation}\label{eq:LJpot}
  W_{{LJ}}(r)= 1 + \left(\frac{a}{r}\right)^{12} -
  2\left(\frac{a}{r}\right)^6 ; \qquad r\equiv|\vec{d}|
\end{equation}
where the normalization has been chosen in such a way that the
equilibrium position is at $r=a$ (the lattice constant) and, for
convenience, with energy minimum zero.

For the on-site potential, we mimic the effect of the 6 oxygen atoms
in each of the two silicate planes above and below the K plane by a
``virtual'' oxygen atom situated in the same position as each K atom:
\[
V(\vec{u}) = \sum_{k=1}^6 V_o(|\vec{u}-\vec{c}_k|)
\]
where $V_o(x)$ is the potential due to each of the 6 fixed virtual
oxygen atoms at lattice positions $\vec{c}_k$, $k=1,\ldots,6$. Since
the lattice modes we have explored are highly anharmonic and thus we
expect large displacements of the potassium atoms, we have chosen a
standard Morse potential for this interaction:
\begin{equation}\label{eq:Morsepot}
  V_o(s) = \frac12 (1-\exp(-s))^2; \qquad s\equiv|\vec{u}-\vec{c}_k|-a
\end{equation}
This form of interaction does not have a hard-core repulsion and would
allow the the potassium atoms to ``pass through'' the virtual oxygens,
which is a better approximation to the on-site effective potential we
are trying to simulate. On the other hand one has to be careful with
the relative scalings between potentials~(\ref{eq:Morsepot}) and
(\ref{eq:LJpot}), otherwise one could dominate the other for large
displacements. It can be checked that for $a\simeq7$ we have an
adequate balance between the two for the typical displacements
involved in the lattice modes studied.

\begin{figure}
\epsfxsize=0.45\textwidth\epsfbox{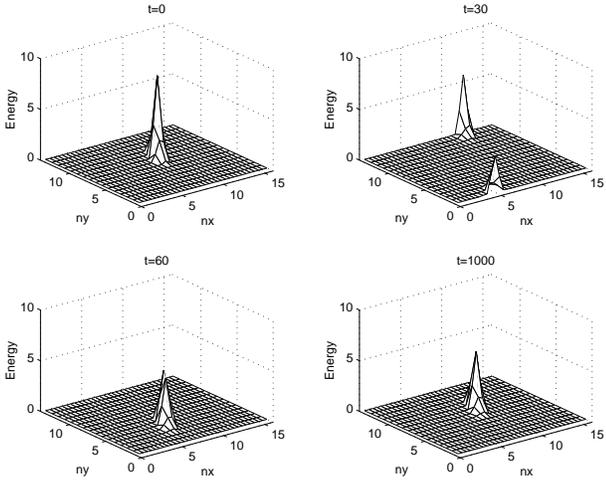}
\caption{Evolution of the energy density profile.}
\label{fig:breather3D}
\end{figure}

We show the result of a typical numerical run in
Fig.~\ref{fig:breather3D}. Simulations are shown on a 16 $\times$ 16
lattice, we have reproduced similar results on larger lattices.  For
the numerical integration we used efficient symplectic
methods~\cite{sc94}.  Initially, we explored different strategies to
obtain moving nonlinear modes in this lattice. It turned out that the
most successful and easy way to generate such solutions was hinted by
the analogue phenomenon of intrinsic localization in 1D nonlinear
lattices. Intrinsic localization of non-moving modes is well-known and
its genericity has been rigorously established. Moving modes are not
so well studied, but they appear generically in simulations of 1D
nonlinear lattices, and in some special cases have been numerically
calculated to high accuracy~\cite{DEFW93}.

Based on experience with 1D lattices, we tried to generate
quasi-one-dimensional moving modes by exciting a few consecutive
atoms, with excitations parallel to their crystal axis. The simulation
shown in Fig.~\ref{fig:breather3D} was obtained simply by exciting
three atoms with a pattern of velocities $-2,4,1$, while the initial
positions remained at their equilibrium points. The rest of the
lattice atoms were at rest.  With this choice of initial conditions
we obtained a remarkably stable moving breather which propagates
along a crystal axis with almost no lateral spreading.  


Fig.~\ref{fig:breather3D}(a) shows the initial distribution of energy,
mainly concentrated on the three initially disturbed atoms (although
some energy now resides in adjacent atoms due to the inter-particle
forces).  The peak moves due to the direction of the imparted
momentum, along the crystal axis away from the position of the viewer.
During the first few time units the peak also modifies its form,
leaving behind some energy, until it has achieved a profile which
apparently moves through the lattice with very little further
distortion.  Fig.~\ref{fig:breather3D}(b) shows the result at $t=30$,
when it is passing through the boundary at the far end of the
rectangle and re-appearing on the near edge (due to the periodic
boundary conditions) which clearly shows the direction of travel.
Fig.~\ref{fig:breather3D}(c) shows the pulse at $t=60$, after it has
traversed the lattice once.  Finally, Fig.~\ref{fig:breather3D}(d)
shows the pulse at a much later time of $t=1000$ when it has traversed
the lattice 12 times, i.e. traveled 192 lattice sites.  Further
integration up to at least $t=7000$ (not shown) shows no appreciable
degredation.  

A close study of the solution shows that most of the energy is
contained at any one time in three or four collinear atoms, with the
surrounding atoms (including those one atomic spacing perpendicular to
the direction of motion) sharing very small amounts of energy.  This
channeling of energy in a very localized way is an interesting
phenomena, we are still studying the details of the mechanism, but it
may be that some sort of discrete self-focusing effect is playing a
part.  Plots (not shown) of the displacements show the characteristic
out-of phase motion of an optical breather~\cite{fl95}.  In the first
few tens of time units the pulse sheds some 10\% of its local energy
to the surrounding lattice, thereafter it adopts a more robust shape
which loses energy at a much lower rate.

Further studies with a variety of initial conditions have shown that
this is a robust phenomena for a range of parameter values.  It has
been observed that the initial velocity perturbations are allowed to
deflect as much as $\pm15^o$ and still produce the moving breather.
However, there are threshold effects: if the initial disturbance is
too weak then the breather will disperse in all directions, which
demonstrates the essential importance of nonlinearity in the QOD
behavior; on the other hand, if it is too strong then the breather
will generate stronger radiation and will eventually become pinned by
the lattice.  It is even possible to generate a breather by giving
just a single atom a nonzero velocity, but in this case the breather
loses more energy to the surrounding atoms before forming itself into
a stable moving structure.  This remarkable behavior supports strongly
the conjecture of quodons (moving breathers) being formed in mica by
energetic cosmic rays, which would produce a large recoil in a single
potassium atom.  From our preliminary studies at least, it seems to be
important that the strengths of the on-site and intersite potentials
be roughly similar.  More details of these effects, and of the effects
of collision between two breathers, will be reported elsewhere.

A straightforward but lengthy calculation can be performed to
calculate the wave speeds of small-amplitude waves in this model.
From this we find that the velocity of the observed ILMs are
sub-sonic, propagating at between 20\% and 60\% of the maximum group
velocity of linear waves.

Although some work has been done on one-component breathers in a
two-dimensional discrete sine-Gordon model~\cite{trp95}, we believe
that this is the first observation of {\em mobile} breathers in a
two-component (i.e.  displacement in both $x-$ and $y-$directions) 2D
discrete lattice.  More work is required to compare the model
potentials used here with the molecular calculations of the full mica
structure.  It would be interesting also to see if similar results can
be found in 2D cubic structures and in full 3D structures.

\subsubsection*{Acknowledgments}  One of us (JLM) would like to thank
the TRACS programme at the Edinburgh Parallel Computing Centre for
support during a three month stay at Heriot-Watt University, during
which time much of this work was done. We also thank J-C Desplat for
his help with visualization software. We are grateful to E.~Abraham,
J.~Blackburn, D.~B.~Duncan, B.~Fleming and Y.~Zolotaryuk for useful
discussions.


\end{document}